\documentclass[preprint]{epl}

\title{Role of mean free path in spatial phase correlation and nodal screening }
\shorttitle{Spatial Phase Correlation \& Mean Free Path}
\author{B.A. van Tiggelen \and D. Anache \and A. Ghysels\footnote{present address: Center for Molecular Modelling, Universiteit Gent, Proeftuinstraat 86, 9000,
Gent, Belgium }} \institute{Laboratoire de Physique et Mod\'{e}lisation des Milieux Condens\'{e}s/CNRS, Maison des Magist\`{e}res/UJF, BP 166,
38042 Grenoble, France.} \shortauthor{B.A. van Tiggelen \etal}


\begin{document}

\maketitle

\begin{abstract}
We study the spatial correlation function of the phase and its derivative, and related, fluctuations of topological charge, in two and three
dimensional random media described by Gaussian statistics. We investigate their dependence on the scattering mean free path.
\end{abstract}

\section{Introduction}
\label{s.intro}

\suppressfloats In disordered media the scattering mean free path (notation $\ell$) denotes the characteristic distance that waves need to
travel to achieve a random phase \cite{frisch}. The length $\ell$ sets the scale in all aspects of multiple scattering.``Mesoscopic" samples
have sizes larger than $\ell$ and smaller than the decoherence length, beyond which the phase is irreversibly destroyed, by either coupling to
environment (quantum) or due to background noise (classical). A third fundamental length, the transport mean free path (notation $\ell^*$),
determines the randomization of the direction of propagation. It is closely related to $\ell$ but the physics is subtly different: whereas
$\ell$ is determined by superposition, $\ell^*$ is also sensitive to interference.

Despite its conceptual importance, $\ell$ is rather unaccessible.
Theoreticians  have found it hard to calculate for dense media
because infinitely many ``Feynman diagrams"
contribute\cite{frisch}, and choices have usually been guided by
effective medium arguments \cite{cpa}. In experiments it is mostly
$\ell^*$ that appears in observables associated with energy
transport \cite{akk}. To get $\ell$ directly we must measure the
`coherent' \emph{field} $\langle\Psi \rangle$, either for
different sample thicknesses or for different times. This requires
a coherent source, phase sensitive detection of an exponentially
small signal, and a good ensemble average. Today, these
requirements can be met for ultrasound \cite{tourin} and
electromagnetic waves \cite{hetero}. A final complication is that
coherent beam experiments always measure the combination of phase
randomization and absorption, quantified by the \emph{extinction}
length $\ell_e$. Here, we investigate the possibility to retrieve
the scattering mean free path of a random medium from the spatial
phase statistics of the scattered waves, rather than from the
coherent field. The phase of a diffuse field can be measured using
a ``passive" source, whose coherence, position and magnitude are
of no relevance, provided it delivers diffuse energy well in
excess of the background noise.

First we introduce the field correlation function in random media
and discuss the role of small absorption. Secondly we discuss the
basic results related to phase statistics and give a rapid
theoretical support. We finally discuss the relation to
topological charge, whose variance has been studied extensively in
literature.


\section{Spatial Field Correlation}
\label{C}
 In a
multiply scattering medium far from the localization threshold, it is widely accepted that the
 complex field obeys circular Gaussian statistics
\cite{Goodman}. Since the
 average field $\langle\Psi (\textbf{r},t)\rangle$ vanishes if the source is far away,
 the only  parameter governing the wave statistics is the
 two-point correlation function
$\langle\Psi (\textbf{r}-\frac{1}{2}\textbf{x},t-\frac{1}{2}\tau ) \Psi (\textbf{r}+\frac{1}{2}\textbf{x},t+\frac{1}{2}\tau)\rangle $
\cite{berry}. We consider a heterogeneous medium in either 2D or 3D, in which all waves suffer from the same small absorption rate $1/ \tau_a$
in energy (absorption length $\ell_a = c\tau_a$).We assume the medium to be large enough to ignore the boundaries.

We consider a wave field $\psi_B(\mathrm{r},t)$ released by a source with spectral density $S(\omega)$. The subscript $B$ refers to a
band-filtering in some frequency band $B $ in which diffusion constant $D$ and absorption time $\tau_a$ are constant ($Bt >1$ is imposed to
resolve the dynamics). Its (ensemble-averaged) spatial correlation at any lapse time $t$ in the coda takes the following diffuse form
\cite{transport},

\begin{eqnarray}\label{corr}
\left<\Psi_B(\textbf{r}-\frac{1}{2}\textbf{x},t ) \Psi_B (\textbf{r}+\frac{1}{2}\textbf{x},t)^* \right> \sim \frac{\exp \left( -\frac{r^2}{4Dt}
- \frac{t}{\tau_a} \right)}{(Dt)^{d/2}}\ \int_B \frac{d\omega}{2\pi }S(\omega)  \textrm{ Im } G(\textbf{x},\omega,\ell_e^{-1} \rightarrow
\ell^{-1} )\nonumber
\end{eqnarray}
This formula shows that, once absorption has been acknowledged by the factor $\exp(-t/\tau_a)$, the field correlation function is proportional
to (the imaginary part of) the \emph{absorption-free} Green function. This is best understood in the frequency domain, where Eq.~(\ref{corr})
gives rise to the following product of exponentials
$$\exp[i(\omega+ i/2\tau_a)(t-\tau/2)] \times \exp[i(\omega+ i/2\tau_a)(t+\tau/2)]^* = \exp(-i\omega\tau)\exp(-t/\tau_a),$$
from which infer that the absorption only appears as a function of the lapse time $t$, and not of the local correlation time $\tau$. A first
observation is thus that in the diffuse time-tail, we can consider the (normalized) spatial correlation of the field to be free of absorption.

In this work we shall ignore the effects of source spectrum and finite bandwidth, which can both straightforwardly be included into Gaussian
statistics \cite{berry2}. If we denote by $C(x)$ the spatial field correlation function at frequency $\omega$ \cite{pnini}, in the vicinity of
some point $\textbf{r}$, deep in the coda at time $t$, and normalized to $C(0)=1$, we find it to be solely dependent on the local Green's
function, a central issue in recent attempts to image passively \cite{retriev}. Contrary to chaotic media, the $C(x)$ of random media always
decays exponentially due to dephasing from scattering. In 3D is $C(x)= \textrm{sinc} (kx) \exp(-x/2\ell)$, whereas in 2D $C(x)= J_0 (kx)
\exp(-x/2\ell)$ \cite{comm}, with $k=2\pi/\lambda$ the wave number. The damped oscillations on the scale of the wavelength $\lambda$ originate
from a superposition of plane waves incident with arbitrary directions but with equal amplitude. It prevents us from seeing the genuine
dephasing that occurs on the \emph{much} longer scale of the scattering mean free path.

\section{Phase correlation}
\label{phasestst} Rather than the field correlation, we propose here to consider the spatial phase correlation $\Xi_B(\textbf{x}, t)
\equiv\langle \Phi(\textbf{r}-\frac{1}{2}\textbf{x},t ) \Phi (\textbf{r}+\frac{1}{2}\textbf{x},t) \rangle$. The phase $\Phi(\textbf{r},t )$ is
defined as the complex phase of the wave function $\Psi_B =A\exp(i\Phi)$.  At any point $\textbf{x}$ the phase is a flat random number between
$-\pi$ and $\pi$. Upon moving to another point $\textbf{x}'$  it will exhibit discontinuous jumps of $2\pi$ between $\pm \pi$. An occasional
jump of a $\pi$ could occur when a singularity with either positive or negative topological charge \cite{freund} is crossed, but such event has
zero probability on a 1D line. Let us now ``unwrap" the phase, i.e correct by hand for the $2\pi$ discontinuities as was done in the frequency
domain \cite{patrickpre} and in the time domain \cite{john}, to get the \emph{unwrapped phase} or cumulative phase $\Phi_U$ that varies
continuously with $\textbf{x}$, and which  in principle can take all values $ [-\infty, \infty]$. This continuation is \emph{not} topologically
invariant, but depends on the exact path chosen to go from $\textbf{r}$ to $\textbf{r}+\textbf{x}$. Gaussian statistics demonstrate that both
the distribution $P(d\Phi/d x)$, and the spatial correlation function $C_\Phi( \Delta x)$ of the phase derivative $d\Phi/dx$ are smooth,
confirming that phase jumps do not affect phase statistics within an infinitesimal increment $dx$. So the integral of $d\Phi/dx$ along the path
should - statistically speaking - be equal to the continuous cumulative phase. If we agree to walk straight along the \emph{x-}axis, this
results in,
\begin{equation}\label{cphase}
    \Xi_U(x) = \int_0^{-x/2} dx' \int_0^{x/2} dx'' \left< \frac{d\Phi}{dx}(x')
      \frac{d\Phi}{dx}(x'')  \right>  = - \int_0^{x/2} dx'\, x' \, [C_\Phi(x') +
      C_\Phi(x-x')].
\end{equation}
The second equality uses that the phase derivative correlation function $C_\Phi(x',x'')$  depends only on $|x'-x''|$

The variance of the unwrapped phase has already been discussed in literature in view of its close connection  with fluctuations in the positions
of zeros of the wave function \cite{freund2,wilkinson,berry2,foltin}. It follows from Eq.~(\ref{cphase}) that,
\begin{equation}\label{cumvar}
   \langle\Phi_U^2(x)\rangle  = 2 \int_0^{x} dx'\, (x-x') \, C_\Phi(x').
\end{equation}

\begin{figure}
    \threefigures[scale=0.44]{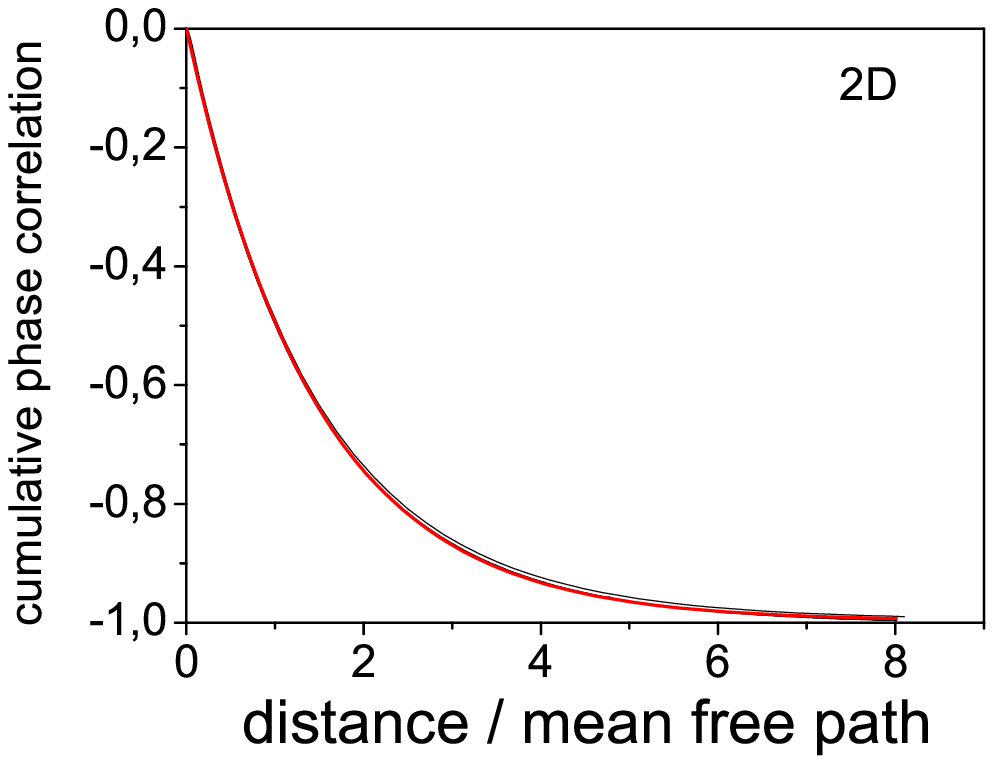}{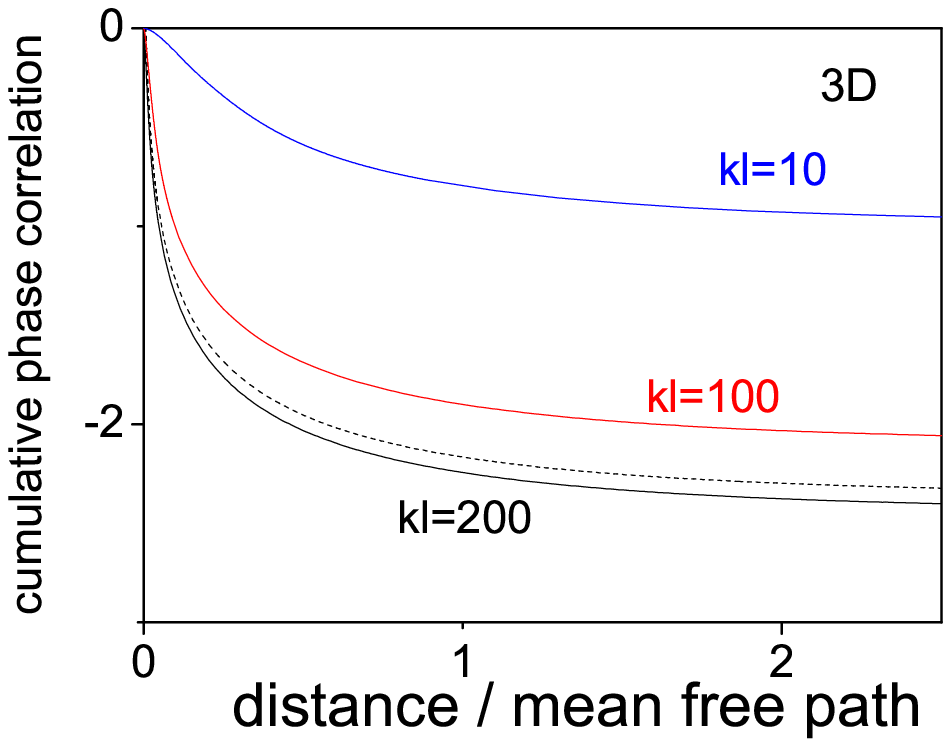}{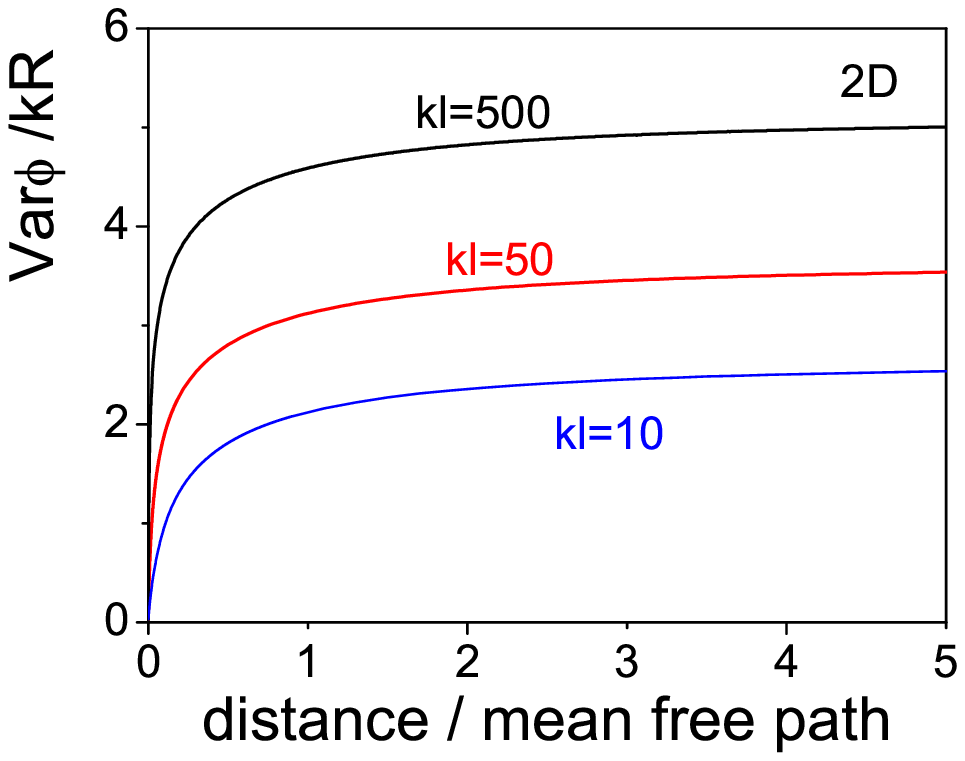}
    \caption{Spatial cumulative (unwrapped) phase correlation $\Xi_U({x})$ versus distance $x$ for a random medium in two dimensions. The distance has been
    measured with respect to the scattering mean free path $\ell$, and the correlation function has been
    normalized to $k\ell/\pi$ which corresponds to (minus) the asymptotic value for large $kR$.
    The curves for $k\ell=10,100,200$ overlap and follow an exponential decay (in red).}
    \label{2Dcorr}
    \caption{Spatial cumulative phase correlation $\Xi_U({x})$ for a random 3D medium, for different values of
    $k\ell$. The asymptotic value varies logarithmically with $k\ell$. Dashed line is obtained by integrating just the
    asymptotic formula~(\ref{dasym}).}
    \label{3Dcorr}
    \caption{Cumulative phase
    variance $\langle\Phi^2_U(x)\rangle $ in 2D random media, divided by the optical path $kx$, against distance $x$, measured in
    units of the mean free path. Two length scales exist. An initial linear rise is visible on the scale of the
    wavelength. Subsequently,
    the phase variance reaches the diffuse plateau after a few mean free
    paths.
    The plateau value depends logarithmically on the value
    $k\ell$. }
    \label{cum}\end{figure}

The phase derivative correlation function $C_\Phi( x)$ can be
calculated from the joint probability function of 4 complex fields
at 4 different positions, which involves both  $C(x)$, $C'(x)$ and
$C''(x)$. The result was earlier reported as an integral
\cite{pre}, but can be put into closed form,
\begin{eqnarray}\label{deri}
C_\Phi(x) &=& \frac{2}{\pi} \int_0^\pi\, d\phi \ \left[-\left(C''+ \frac{C(C')^2}{1-C^2} \right) \cos\phi H(C\cos\phi)
+\frac{(C')^2}{1-C^2}\sin^2\phi H'(C\cos\phi)\right] \nonumber
\\&=& \frac{1}{2} (\log C )'' \log (1-C^2)
\end{eqnarray}
with $H(y)\equiv \arctan(\sqrt{1+y}/\sqrt{1-y})/\sqrt{1-y^2}$.
Already for $x>\lambda$, C is small enough to approximate $C_\Phi(
x)$ by its asymptotic limit, and we find
\begin{eqnarray}\label{dasym}
C_\Phi(x>\lambda) \rightarrow \frac{1}{2}[ (C')^2 -C''C ] = \left\{\begin{array}{c}
    ({k}/{\pi x} ) \exp(-x/\ell)\ \  \ (2D) \\
({1}/{2x^2} ) \exp(-x/\ell)\ \ \ (3D) \\
\end{array} \right.
\end{eqnarray}
We see that, contrary to the field correlation $C$, $C_\Phi$ exhibits no oscillations on the scale of the wavelength. Equation~(\ref{dasym})
shows that the mean free path $\ell$ assures convergence of Eq.~(\ref{cphase}) in both dimensions, making $\ell$ the characteristic length scale
in spatial phase correlation. In fig. 1 we plot $\Xi_U({x})$ as a function of $x/\ell$ for 2D. For all values of $k\ell$ it follows a universal
exponential decay towards the asymptotic value $-k\ell/\pi$. In 3D the asymptotic value varies only logarithmically with $k\ell$ in 3D (fig. 2)
and two characteristic length scales appear, $\lambda$ and $\ell$. A measurement of $\Xi_U({x})$ would thus give us direct access to $k\ell$ and
$\ell$ in both cases.

The same study has been performed for the phase variance. We infer
from Eq.~(\ref{cumvar}) that the parameter
$\langle\Phi^2_U(x)\rangle/x $ is less sensitive to large values
for $x$ than the correlation $\Xi_U({x})$. In 2D random media
(fig. 3), its plateau value $D_\Phi $ (the ``phase diffusion
constant") depends logarithmically on $k\ell$, although the
diffuse plateau is still reached  after typically one mean free
path. In 3D (dashed line in fig.~4) we see that after an
approximate linear rise, $\langle\Phi^2_U(x)\rangle/x $ converges
already after a few wavelengths to the asymptote $D_\Phi = 1.922
\times k $, independent of $k\ell$. We conclude that the
cumulative phase correlation is sensitive to the mean free path in
both 2D and 3D, unlike the phase variance which depends on $\ell$
only in 2D.

\section{Statistics of Topological Charge}
\label{stattopo} It was realized recently by Wilkinson
\cite{wilkinson} that the spatial phase correlation function is a
key element in a old discussion on the statistics of topological
charge of Gaussian field inside a closed contour. This discussion
goes back to Halperin \cite{halperin} and Nye and Berry
\cite{berry}, with significant contributions later by Berry and
Dennis\cite{berry2}, and Freund and Wilkinson \cite{freund2}. The
zero's of a complex function in $d$ dimensions are located on
$d-2$ ``nodal" hypersurfaces, nodal lines in 3D and nodal points
in 2D. The topological charge $Q$ enclosed by a 2D surface is
defined as the number of nodal points weighed by their topological
sign. This number is determined by the order of the vortex
surrounding the nodal point \cite{freund}. For complex Gaussian
random fields topological charges different from $\pm 1$ are
highly improbable \cite{berry}. The following relation holds,

\begin{equation}\label{topo}
    \oint_\Gamma d\textbf{r} \cdot\nabla\Phi(\textbf{r})= 2\pi Q
\end{equation}
with $\Gamma$ the line contour enclosing the surface. This
relation is reminiscent of the quantized rotation of superfluids,
and is deeply related to the Rouch\'e theorem in complex analysis.
Its validity can be understood by applying Stokes' theorem to the
function $\nabla \log \psi(\textbf{r})$, whose imaginary part has
been analytically continued  along the contour.

\begin{figure}
    \threefigures[scale=0.41]{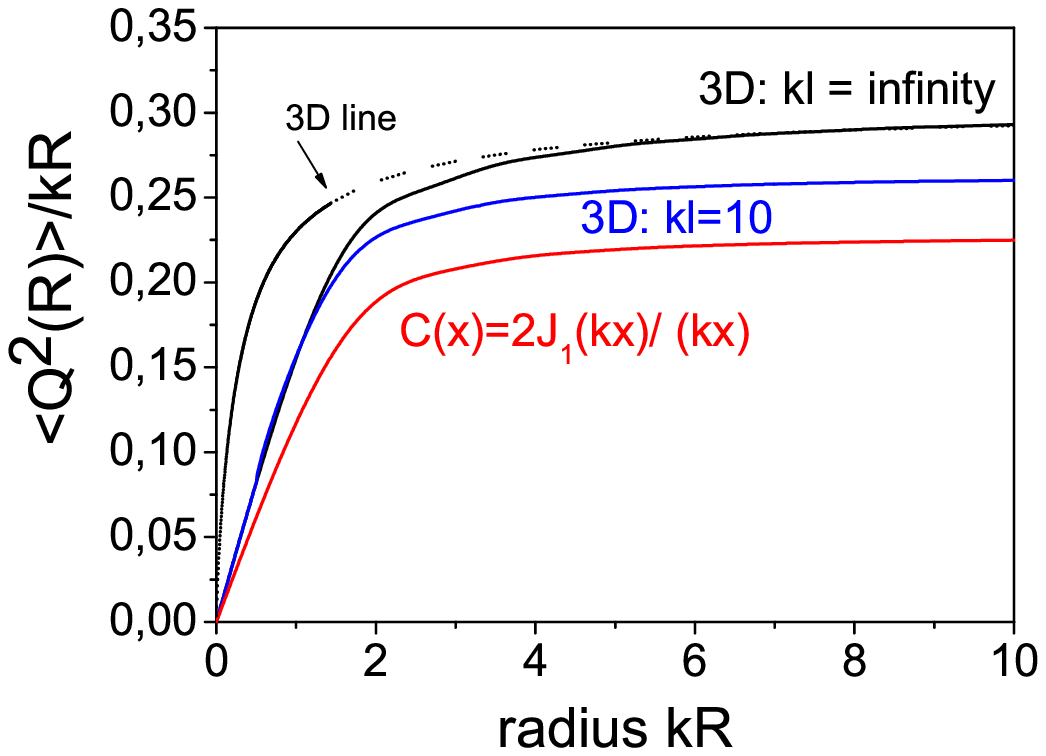}{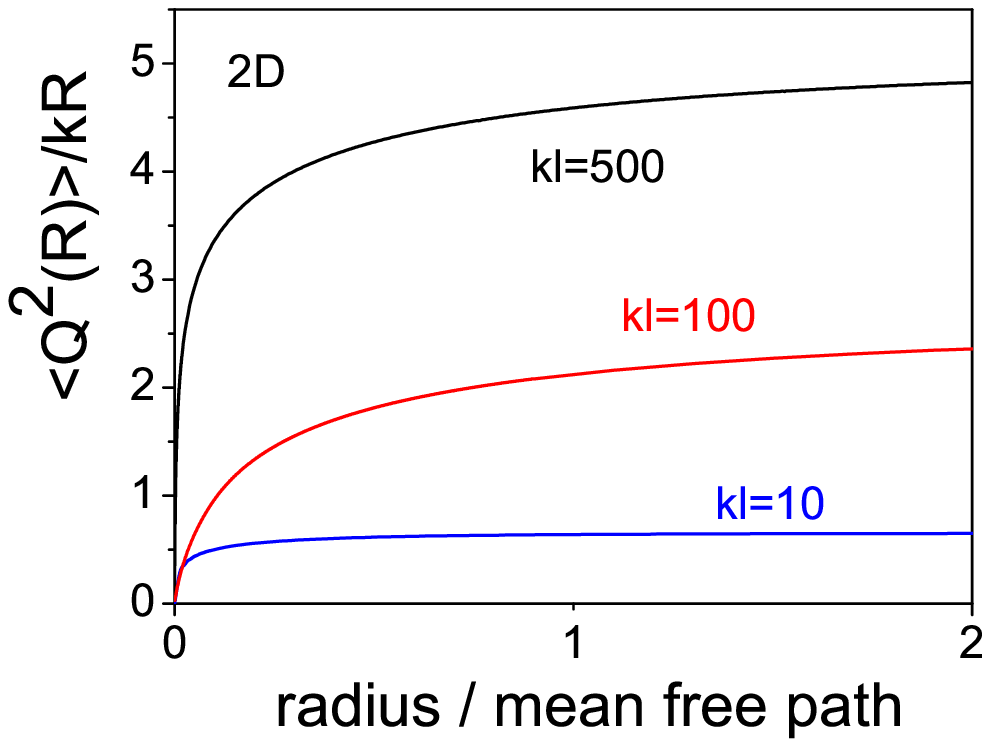}{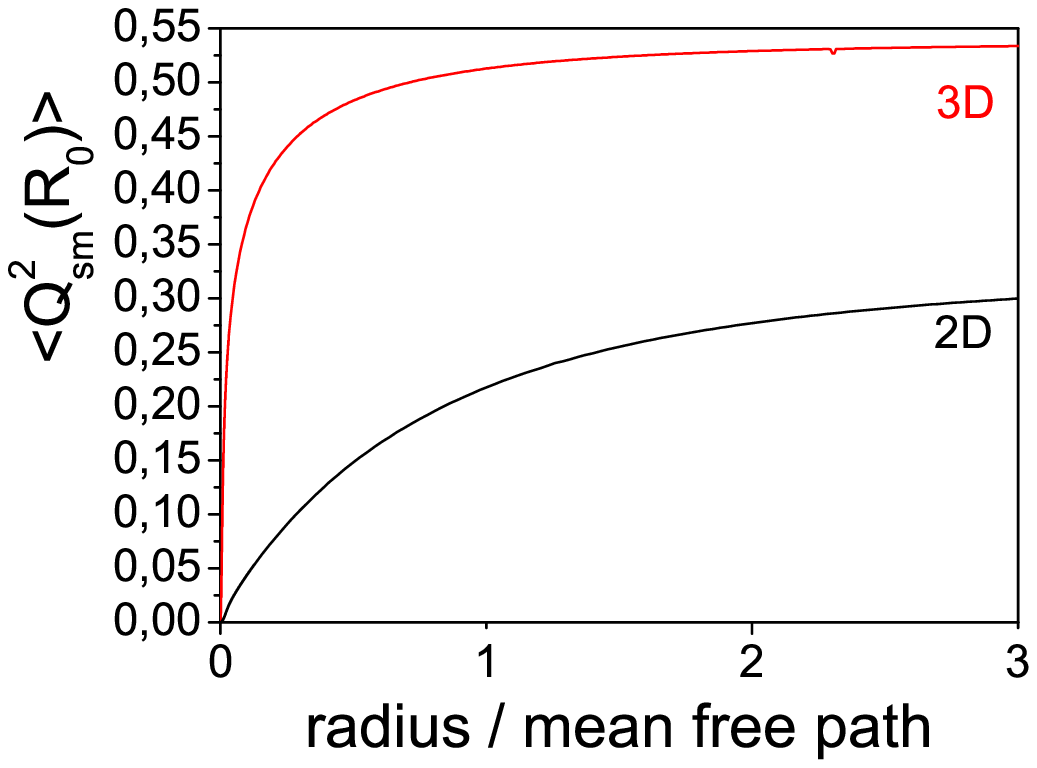}
    \caption{Fluctuations of the topological charge accumulated inside a circle of radius $R$ of a Gaussian random field.
    On the vertical axis is plotted the variance of $\langle Q^2\rangle $ divided by  $2\pi
    R/\lambda$, i.e. the circumference in units of the wavelength. Upper two solid lines: 3D random medium with $k\ell =\infty , 10$. Dashed
    curve is the cumulative phase variance - divided by $2\pi$, see Eq.~(\ref{topo}) - obtained by walking along a straight line
    with length $2\pi R$. The bottom solid curve is calculated for a field correlation
    function $C(x)=2J_1(kx)/kx$ \cite{freund}. In all cases the characteristic length scale is the wavelength, and not the mean free path}
    \label{contour3D}
    \caption{Fluctuations of the topological charge in 2D, accumulated inside a circle of radius $R$.
    On the vertical axis is plotted the variance of $\langle Q^2\rangle $ divided by the circumference $2\pi R/\lambda$
    in units of the wavelength.
    On the horizontal axis the radius of the circle is normalized by the scattering mean free path.
    The characteristic length scale in 2D is the mean free path. The variance of the topological charge depends
     logarithmically on the value $k\ell$. }
    \label{contour2D}
    \caption{Variance of the smoothed topological charge in 3D (normalized to $k\ell$) and in 2D
    (normalized to $\log k\ell$). The
    smoothing is done on $Q(R)$ with the Gaussian
    $\exp(-R^2/ R_0^2)$, as a function of $R_0$ in units of the scattering mean free path. For $R_0>\ell$ it
    converges to a constant number dependent on dimension.}
    \label{contoursmooth}
\end{figure}

Equation~(\ref{topo}) facilitates a study of the statistics of the
topological charge $Q$. In the diffuse regime the phase gradient
in any direction vanishes on average, so $<Q>=0$. Following
Wilkinson \cite{wilkinson} we will relate the variance of $Q$ to
the correlation of the phase derivative. More precisely, if
$\theta \times R $ measures the length along a circular contour
with radius $R$ we find,
\begin{eqnarray}\label{circle}
    \left<Q^2(R)\right> &=& \frac{1}{(2\pi)^2} \int_0^{2\pi} d\theta \int_0^{2\pi}
    d\theta'  C_\Phi(\Delta \theta)  =
    \frac{1}{2\pi} \int_0^{2\pi} d\Delta\theta  \,
      C_\Phi(\Delta \theta)
\end{eqnarray}
The phase derivative correlation $ C_\Phi(\Delta \theta) \equiv
\left<
\partial_{\theta} \Phi(\theta) \partial_{\theta} \Phi(\theta ')\right>$ is, \emph{mutadis-mutandis}, given by
Eq.~(\ref{deri})  (with $(') =\partial_{\Delta\theta }$ and $C=
C[2kR\sin(\Delta\theta/2)]$). One basic feature of
$\left<Q^2(R)\right>$ is already known. If we would assume all
nodal points $n$ to have random charges $Q_n=\pm 1$,
\emph{independent} of each other, we would find $\left<(\sum
Q_n)^2 \right> \sim R^2$, i.e.  proportional to the surface.
Actually, they are not independent but tend to be perfectly
screened \cite{halperin,berry2}, which affects the quadratic law.
We shall define the screening length $\xi$ as the typical length
beyond which the asymptotic form of the charge variance is
reached. Wilkinson and Freund report a linear, ``diffuse"
asymptotic form \cite{freund},
\begin{eqnarray}\label{Qfreund}
     \lim_{R\rightarrow \infty} \frac{\left<Q^2(R)\right>}{R}  =
     \frac{1}{\pi}\int_0^\infty dx \, \frac{C'(x)^2}{1-C(x)^2}
\end{eqnarray}
We have verified that Eq.~(\ref{circle}) is consistent with this diffusion law, with the same phase diffusion constant obtained earlier for the
phase accumulated along a straight line (Fig.~3 and 4). Berry and Dennis \cite{berry2} calculate $\left<Q^2(R)\right>$ from the topological
charge correlation function, which is, due to screening, sensitive to ``edge effects". When the charge distribution is smoothed with the
Gaussian $\exp(-R^2/R_0^2)$ (with smoothed surface $\pi R_0^2$), the quadratic law valid for small $R_0$ turns over into the asymptote,
\begin{eqnarray}\label{Qberry}
     \lim_{R_0\rightarrow\infty} \left<Q_{\mathrm{sm}}^2(R_0)\right>
     =\int_0^\infty dx \, x \, \frac{C'(x)^2}{1-C(x)^2}
\end{eqnarray}
provided the integral converges. In figs. 4 and 5 we have calculated $\left<Q^2\right>$ for a disk in 2D and 3D random media. For small $kR$ we
see that $\left<Q^2\right> \sim k^2 R^2$ for both dimensions. In 3D $\left<Q^2\right>$ at large $R$ depends very weakly on the mean free path
$\ell$ since the integral in Eq.~(\ref{Qfreund}) converges even for $\ell =\infty$. For the field correlation $C(x)= 2J_1(kx)/kx$ \cite{freund}
the curve is very similar though with a somewhat smaller phase diffusion constant. In both cases this yields a screening length proportional to
the wavelength. For 2D random media we can see that $\left<Q^2\right>$ depends logarithmically on $k\ell$. The topological screening length now
depends on the mean free path. The variance of \emph{smoothed} charge  $\left<Q_{\mathrm{sm}}^2\right> $ however (fig.~6), converges to
$k\ell/\pi $ in 2D and to $ 0.54 \, \log k\ell$ in 3D, with a screening length typically equal to the $\ell$. Note that in 2D chaotic media, for
which $C(x)=J_0(kx)$, the integrals~(\ref{Qfreund}) and (\ref{Qberry}) would diverge and the topological charge variance would actually increase
as $\left<Q^2\right>\propto kR \log(kR) $, i.e. is slightly superdiffuse, whereas its smoothed version $\left<Q^2_{\mathrm{sm}}\right>$ no
longer reaches a constant but becomes diffuse: $\left<Q_{\mathrm{sm}}^2\right> \rightarrow kR_0/(2\sqrt{\pi}) $.

\section{Conclusions}We have shown that in the late coda of waves
propagating in 2D and 3D random media, the scattering mean free path governs the spatial fluctuations of both the phase derivative $d\Phi/dx$,
cumulative phase and the topological charge, measured inside a surface of size $A$. Screening makes topological charge fluctuations grow slower
than $A$. The screening length is defined as the typical length beyond which they reach their final behavior. In 2D it is proportional to the
mean free path, in 3D to the wavelength. If the charge is smoothed with a Gaussian the screening length is typically equal to the mean free
path, in both dimensions. This highlights the subtle role of smoothing screened charge.

 The spatial correlation of
$d\Phi/dx$, whose measurement requires only four phase-sensitive
detectors, decays exponentially with the scattering mean free path
$\ell$ as the sole length scale. In seismic measurements, often
dominated by 2D Rayleigh waves, this may be a novel opportunity to
measure the scattering mean free path of surface waves. Such
measurement would not be sensitive to absorption, neither to the
precise source location, quite opposed to a measurement of the
coherent beam.

\acknowledgments The authors are indebted  to L. Margerin, A. Genack, R. Weaver, J. Page and M. Dennis for useful discussions.


\begin{thebibliography}{0}
\bibitem{comm} One can show that
$C'(0)=0$ and $C''(0) = -\frac{1}{d}k^2c_0/v_E$ with $v_E$ the transport velocity, inconsistent with the exponential decay proposed for $C(x)$,
thus only correct for large $x$. In this work we exclude the exponential factor for $x < \lambda$, and also ignore phase accumulation inside the
scatterers, which is the reason why $v_E$ should appear.
\end{thebibliography}
\end{document}